\documentclass[%
 aip,
 jmp,%
 amsmath,amssymb,
preprint,
]{revtex4-1}

\usepackage{graphicx}
\usepackage{epstopdf}
\usepackage{dcolumn}
\usepackage{bm}
\usepackage{epsfig}

\begin{document}

\preprint{AIP/123-QED}

\title{Visualization of the flow profile inside a thinning filament during capillary breakup of a polymer solution via particle image velocimetry (PIV) and particle tracking velocimetry (PTV)}

\author{S. Gier}
 \author{C. Wagner}
  \email{c.wagner@mx.uni-saarland.de}
 \homepage{http://www.uni-saarland.de/fak7/wagner/}
 \affiliation{$^1$Technische Physik, Universit\"at des Saarlandes, Postfach 151150, 66041
Saarbr\"ucken, Germany}

\date{\today}

\begin{abstract}

We investigated the flow profile of a polymer solution in a thinning capillary bridge. Fluorescent tracer particles with a diameter of 3$\mu$m were used to visualize the flow. The cylindrical shape of the filament introduced strong optical abberations that could be corrected for, and we were able to characterize the flow in filaments with a thickness ranging from 150 to 30 $\mu$m. In the first regime when the filament was still sufficiently large, we used a PIV algorithm to deduce the flow field. At later stages when the number of particles in the observation plane decreased a PTV algorithm was used. The main two results of our measurements are as follows. First, the flow profile at the formation of the cylindrical filament is highly inhomogeneous and there is only flow in the outer parts of the filament. Second, we find that in most parts of the regime, where the temporal radius of the thinning filament can be fitted with an exponential law the flow indeed is purely extensional.
\end{abstract}




\maketitle

%

\section{Introduction} \label{sec1}
\begin{figure}[h!]
  \includegraphics[width=0.75\linewidth]{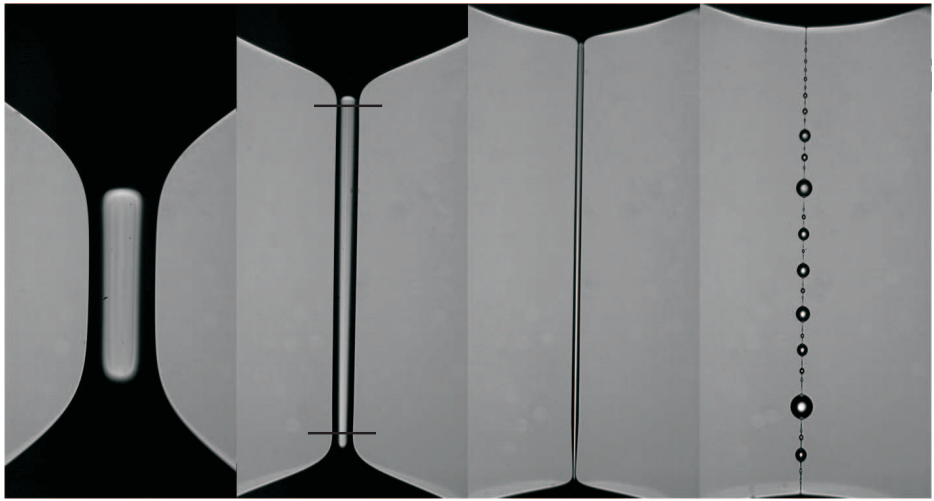}\centering
  \caption{Shadowgraph images from the capillary breakup of a $0.2$ weight percent polyethylene oxide (PEO, Mw= $4 \times 10^6$ g/mol) in a 60/40 weight percent glycerol/water solution. Images are taken at t = -0.05, 0.25, 0.43 and 0.65 s (cf. Fig. \ref{Durchmesser_Kap7}). The size of the images is 0.55 $\times$ 1.1 mm. The horizontal lines in the second image indicate the region shown in Fig. \ref{PIV-Bilder}. The last image shows the final instability of the viscoelastic thread when many small droplets are formed.}\label{PEO_Beispiel}
\end{figure}

The Rayleigh-Plateau instability is the classical example for hydrodynamic instabilities in free surface flow \cite{E97}. It describes the disintegration of a liquid column into separate droplets via initial  exponential growth of the sinusoidal deformation of the free surface. After sufficient viscous stretching, the system evolves into a nonlinear regime, and close to pinching, the system does not depend anymore on the primary configurations. The break-up process is described by self-similarity solutions. The addition of a small amount of polymers does not significantly alter the primary, linear instability because the solution has a very low linear elastic contribution and the polymers do not yet affect the flow at low shear rates \cite{Wagner2005}; only, when the (elongational) flow rates become large enough the polymers intervene the flow. The thinning process is strongly slowed down, and instead of further necking of the capillary bridge, a cylindrical filament is formed that thins exponentially in time. The flow is supposed to be purely extensional. Figure \ref{PEO_Beispiel}, for example, shows the capillary breakup of a polymer solution.

Extensional flow is of primary interest for the characterization of polymers because elastic stresses can build up that are orders of magnitude larger compared to simple shear flow. The experimental determination of the elongational viscosity is still a challenging task and the characterization of this cylindrical thread is one of the rare reliable methods to extract at least some information on the elongational properties of polymer solutions. This experimental setup is also referred to as CaBER (capillary break-up extensional rheometer) and a commercial version is available as well. In spite of numerous experimental and theoretical investigations, there is still no complete understanding of the capillary break-up process \cite{Bazilevskii1981,Renardy94,Chang1999,Stelter2000,Anna2001,Li2003,Wagner2005,Etienne2006,Clasen06,Tirtaatmadja06,Morrision2010,Zell2010,Bhat2010}. Most theories indeed predict an exponential thinning of the cylindrical thread \cite{Anna2001}, but the determination of the correct time scales remains an open issue \cite{Clasen06}. Furthermore, the elongational viscosity can be only deduced once a parallel filament has formed, but at this moment in time, the polymers have already accumulated a considerable amount of stress, and it is not clear how the transition from the primary thinning to the cylindrical filament shall be described. Numerical simulations indicate that there is a thin layer of large stress near the free surface of the thinning capillary bridge, while our experiments show the existence of a small layer with flow close to the free surface or no flow at all in the inner parts of the cylinder near the stagnation point where the flow is mostly elongational. Experimental observations also show that the flow becomes very inhomogeneous at the final stage of the thinning process where the filament becomes unstable to the formation of secondary droplets and even some flow-induced phase separation has been reported \cite{Sattler2008}. However, our measurements show that in the intermediate regime, the flow indeed is purely elongational, and the only irregularity is the movement of the stagnation point.

The experimental characterization of the flow field in a thinning capillary bridge implies certain difficulties due to the small sizes that even change in time and the optical aberrations due to the cylindrical form of the filament. In microscopic flows, fluorescent particles are often used to visualize the flow. Even if they are illuminated along the optical axes, the exiting light can be filtered before the camera, and only the fluorescence signal is observed that leads to a good contrast. The small sizes allow the use of microscopic objectives with a large numerical aperture. The desired plane of observation is determined by the field of depth, and there is no need to shape the incident light beam, e.g., in the form of a sheet \cite{Meinhart99,Mielnik04}.

\section{Theoretical Background} \label{sec2}

\subsection{Capillary breakup of a liquid bridge or droplet detachment}
The classical Rayleigh-Plateau instability describes the instability of a cylinder of infinite length, but the theoretical growth rate has been found to be in very good agreement with experimental data from a capillary thinning bridge \cite{Rothert01,Wagner2005}. For an infinite thread, the instability evolves spatially in the form of a sinusoidal deformation, and for a short capillary bridge, the necking evolves in a symmetric manner. In the middle of the necking capillary bridge, a stagnation point is formed around which the flow is elongational. This elongational flow is very efficient in stretching the polymers, and when the flow rate times the polymer relaxation time, i.e. the Weissenberg number, is $Wi=\dot\epsilon \lambda>0.5$, the polymers intervene the flow. For such a flow, the Oldroyd-B model predicts a temporal evolution of the minimum filament diameter in the form:
\begin{equation}
h(t)=h_0e^{-t/\left(3\lambda_O\right)}. \label{Filament_Polymer}
\end{equation}
$h_0$ is the diameter of the filament at the point when
exponential thinning starts, and $\lambda_O$ is the polymeric relaxation time. The elongational viscosity can be deduced from the measurement of $h(t)$ by the force balance of the capillary pressure and the polymeric stress. It follows that

\begin{equation}
\eta_e=\frac{2\sigma}{h(t)\dot{\epsilon}}. \label{elongational_viscosity}
\end{equation}

$\dot{\varepsilon}$ is the elongation rate, and $\sigma$ the surface tension.
A pure elongational flow in cylinder coordinates is defined by a velocity in axial direction and in radial direction in the form
\begin{equation}
v_z=\dot{\varepsilon}z \label{axial}
\end{equation}
and
\begin{equation}
v_r=-\frac{1}{2}\dot{\varepsilon}r. \label{radial}
\end{equation}

$z$ and $r$ are the axial and the radial position on the filament, respectively. Thus, the elongational rate can be determined by simply measuring the diameter $h(t)$ of the filament
\begin{equation}
\dot{\varepsilon}(t)=-2\frac{\partial_t h(t)}{h(t)}.
\label{Dehnrate1}
\end{equation}

Together with Equ. \ref{Filament_Polymer} one finally gets an
elongation rate which is constant in time:
\begin{equation}
\dot{\varepsilon}(t)=\frac{2}{3 \lambda_O}=\dot{\varepsilon}.
\label{Dehnrate2}
\end{equation}

The exponential regime might last until the polymers are fully elongated.
When the polymers inside the filament are completely stretched, the
liquid might behave as a Newtonian fluid, but with a strongly increased
viscosity. In this regime, the diameter of the filament is supposed to decrease
linearly in time according to the self similarity solutions for viscous liquids \cite{E97}.

\section{Experimental methods} \label{sec3}

\subsection{Sample preparation}

The polymer in our study was polyethyleneoxide (PEO) with a molecular weight of $M_W=4\times10^6$ g/mol (Sigma-Aldrich, Munich, Germany). PEO is a highly flexible polymer that has often been investigated in elongational measurements \cite{Wagner2005,Tirtaatmadja06,Zell2010,Sattler2008,Amarouchene01,CooperWhite02,Lindner03,Rodd05}. The polymer was dissolved in a $60/40$-weight percent glycerol-water solution at a concentration of  $c = 0.2$ weight percent, and the overlap concentration was $c_{ov}=0.07$ weight percent. Both the semidilute concentration \cite{Yao00, Rothstein02,McKinley01} and the viscosity of the solvent were chosen such that the characteristic time scales were slow enough to allow for a quantitative measurement of the flow profile. In general glycerol is hygroscopic, but this is only relevant at very concentrated glycerol solutions. The solution was prepared with the following protocol: the polymer was slowly poured in water first. After some minutes of swelling at rest, the polymer-water mixture was gently stirred for 24 hours. After that, the glycerol was added, and the solution was stirred gently again for 24 hours. The solution was measured within a few hours after preparation to minimize degradation of the polymers. This protocol lead to very reproducible results that were in agreement with data from the literature \cite{Amarouchene01,CooperWhite02}. For the PIV and PTV measurements, a small amount of fluorescent particles (0.6 $wt\%$) was added to the solution just before the measurement. The tracer particles had a diameter of 3$\mu$m, and their absorption maximum was at a wavelength of 530 nm while the emission maximum was at 630 nm.  The samples were characterized with a Thermo Haake MARS rheometer using cone plate geometries. The viscosity at a shear rate of $\dot{\gamma}=1$ was $\eta_{1}\approx 70mPas$, and shear thinning was present up to a shear rate of $\dot{\gamma}\approx 300$  with $\eta_{300}\approx 40mPas$. Above this shear rate, the fluid became unstable to a viscoelastic instability. The surface tension of the solvent was $\gamma\sim65mN/m$ and its density $\rho=1150 kg/m^3$. The addition of the polymer did not change the surface tension within the measurement resolution of the pendant drop method $\pm 2mN/m$. We should not expect that additional Marangoni stresses at the free surface of the viscoelastic filament affect the flow because the heat absorption of the laser was very low ($< 1^\circ C$). We also tried to work with less concentrated or less viscous solutions but the dynamic was always to fast to allow for a quantitative PIV analysis.

\subsection{Experimental setup}

Our self-built Capillary Breakup Extensional Rheometer (CaBER) consisted of two circular stainless steel plates with a diameter of 2 mm. A droplet of the polymer solution was put between the two plates. The upper plate was then moved away slowly with a speed of 0.05mm/s by a linear motor that was controlled by a PC. When the distance of the two plates overcame the critical length for Rayleigh-Plateau instability, the capillary bridge started to thin. This experimental protocol differs from the standard procedure where a steep step is performed to separate the plates and to introduce a kind of well-defined pre-stretch but similar to the slow retraction method presented in Ref. \cite{Miller09}. We find that our protocol is first more closely related to the more interesting case of a capillary bridge that thins due the Rayleigh-Plateau instability, and second, we find that the thinning of the filament is much more stable, e.g., there are less oscillations in the filament that disturb the PIV-measurements significantly. For the simple shadowgraph measurements (cf. figure \ref{PEO_Beispiel}), the thinning bridge was filmed with a high-speed camera that was equipped with a microscope objective with $5 \times$ magnification.

\begin{figure}[h!]
  \includegraphics[width=0.75\linewidth]{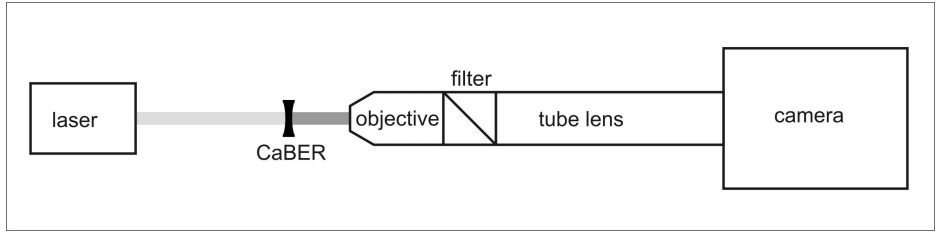}\centering
  \caption{Sketch of the experimental setup.}\label{Transmission}
\end{figure}

For the flow visualization measurements, the  fluorescent tracer particles in the solution could be illuminated either with a continuous Nd:YAG laser ($\lambda$=532nm, $P_{max}\approx$250mW) or with a pulsed Nd:YLF laser ($\lambda$=527nm, $P_{max}\approx$16W). We chose a transmission geometry, and the part of the laser light that was not diffracted by the filament passed through the sample before entering the $20 \times $ magnification microscope objective with an numerical aperture of $NA =0.4$ (Fig.\ref{Transmission}). A filter cube that was located between the objective and the tube lens (see Fig. \ref{Transmission}) contained a 605/55nm band-pass filter that protected the camera chip from the laser radiation, but let pass the fluorescent light.

\begin{figure}[h!]
  \includegraphics[width=0.65\linewidth]{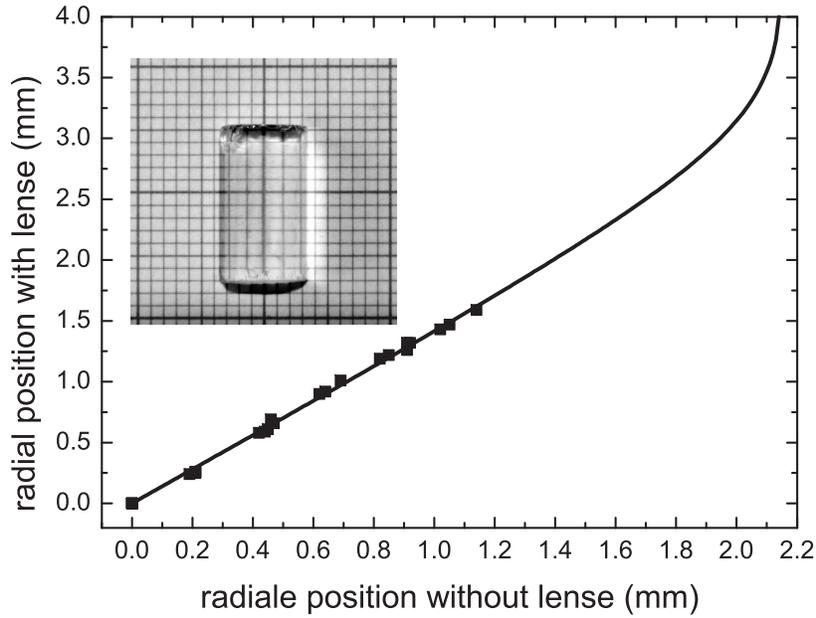}\centering
  \caption{Inset: Cylindrical lens on millimeter paper. Squares: Measured radial positions of the vertical lines recorded with and without the cylindrical lens on the millimeter paper. The line is given by Equ. \ref{Deltax}.}\label{Zylinderlinse}
\end{figure}

Due to the cylindrical shape of the filament, the image of the fluorescent tracer particles is distorted. We only measured in the middle plane of the filament, and the effect can be visualized with a half-cylinder lens. Figure \ref{Zylinderlinse} shows a half cylinder lens lying on a sheet of millimeter paper. Plotting the radial positions of the lines recorded via a cylindrical lens versus those without a lens, one gets a straight line, at least if one is not too close to the radius r. This is in good agreement with a theoretical consideration based on simple geometrical optics. A point source at a distance $x$ from the center of the lens perpendicular to the optical axes seems to be shifted by a length $\Delta x$ that is given by

\begin{equation}
\Delta x=r cos\left(arcsin\left(\frac{x}{r}\right)\right) tan\left(arcsin\left(n_2\frac{x}{r}\right)arcsin\left(\frac{x}{r}\right)\right). \label{Deltax}
\end{equation}

$n_2$ is the refractive index of the cylinder lens. Equ. \ref{Deltax} predicts a close to linear dependency of $\Delta x (x)$ if x is smaller than $x\gg r/n_2$. This relation allows for a simple correction of the geometrical aberrations. However, we were still limited to the regime in which the parallel filament already had been formed. Before the formation of the parallel filament, there is an additional curvature in the direction of the longitudinal axes that is more difficult to correct for.

In the course of thinning of the filament, the number of observable particles was not constant, and we had to start with a comparably high concentration of particles of 0.6 weight percent. A PIV analysis was only possible at the beginning of the thinning process. During the breakup, the number of observable particles decreased, and a PTV analysis was more reasonable from the moment when single particles were distinguishable. Devasenathipathy et al. \cite{Devasenathipathy03} used $\mu$PIV as well as PTV to characterize the flow profile at the intersection of a cross-channel. Their measurements showed that the two techniques provide the same results. This was also confirmed by simulations and is in accordance with our findings.

The images were captured at a rate of up to $150$ Hz and afterwards analyzed by the MatPIV algorithm \cite{MatPIV}. In our measurement, a long but thin filament was investigated, and the interrogation window width was set to 16 pixels in the radial direction and 64 pixels in the axial direction. The overlap of the windows was 50$\%$ for both directions. The depth of the interrogation plane $\delta$ was determined by the numerical aperture of our microscope objective, and for the $3 \mu m$ particles we get $\delta = 2.4 \mu m$\cite{Meinhart99}. The interface between the areas with and without particles moves in the radial direction because the filament thins in time. This movement might lead to spurious velocity vectors in the analysis such that non-vanishing velocity components outside the filament can appear that have to be sorted out manually or by appropriate filtering. For the PTV analysis the centroid algorithm from Lab-View (National Instruments, Austin, Texas) was used.

The smallest capillary threads that were evaluated had a diameter of $h\approx30 \mu m$ and the concentration of the microbeads were low enough to not expect any alteration of the flow due their presence, similar to what is found in standard microfluidic experiments \cite{Meinhart99}. Furthermore, we did not observe any concentration enhancement of the microbeads near the surface of the viscoelastic thread and we assume that the surface tension is not significantly affected by the microbeads.

\section{Measurements and Results} \label{sec4}

\begin{figure}[h!]
\includegraphics[width=0.75\linewidth]{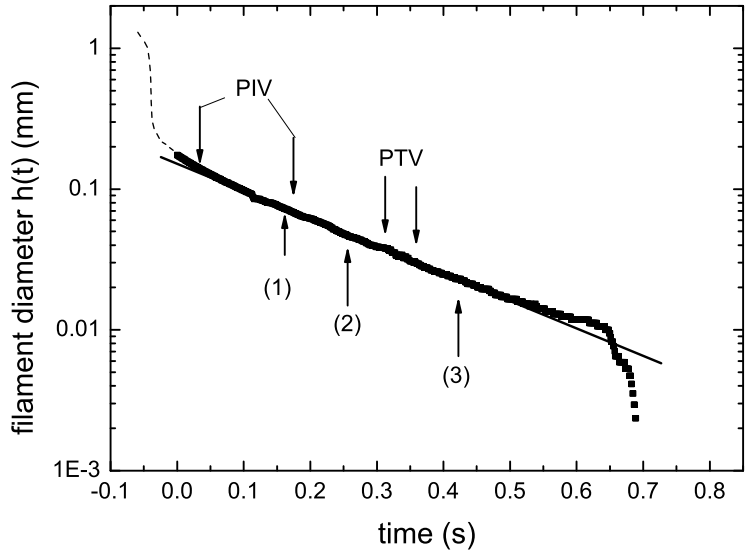}\centering
\caption{Filament diameter versus time. The dashed line is taken from the simple shadowgraph images. It shows the beginning of the capillary instability before the parallel filament has formed. The squares are taken from the background signal of the PIV images (Fig. \ref{PIV-Bilder}). The straight line depicts an exponential fit to the experimental data in the range $0.1<t<0.5$. The characteristic time scale $\lambda_C = 227 $ms is calculated according to Equ. \ref{Filament_Polymer}, and the elongation rate is $\dot{\varepsilon} = 8.81s^{-1}$ . The black arrows below the curve (ticketed with (1), (2) and (3)) mark the moments at which the pictures in Fig.  \ref{PIV-Bilder} are captured. The two left arrows above the curve (PIV) label the transitions in the PIV analysis, which are shown in Fig. \ref{PIV_Anfang} and in Fig. \ref{PIV-Profile_Ausschnitt}. The two right arrows above the curve (PTV) locate the temporal borders in between which the PTV analysis shown in Fig. \ref{PTV_alle} is performed.}\label{Durchmesser_Kap7}
\end{figure}

\subsection{PIV analysis}
\begin{figure}[h!]
  \includegraphics[width=0.75\linewidth]{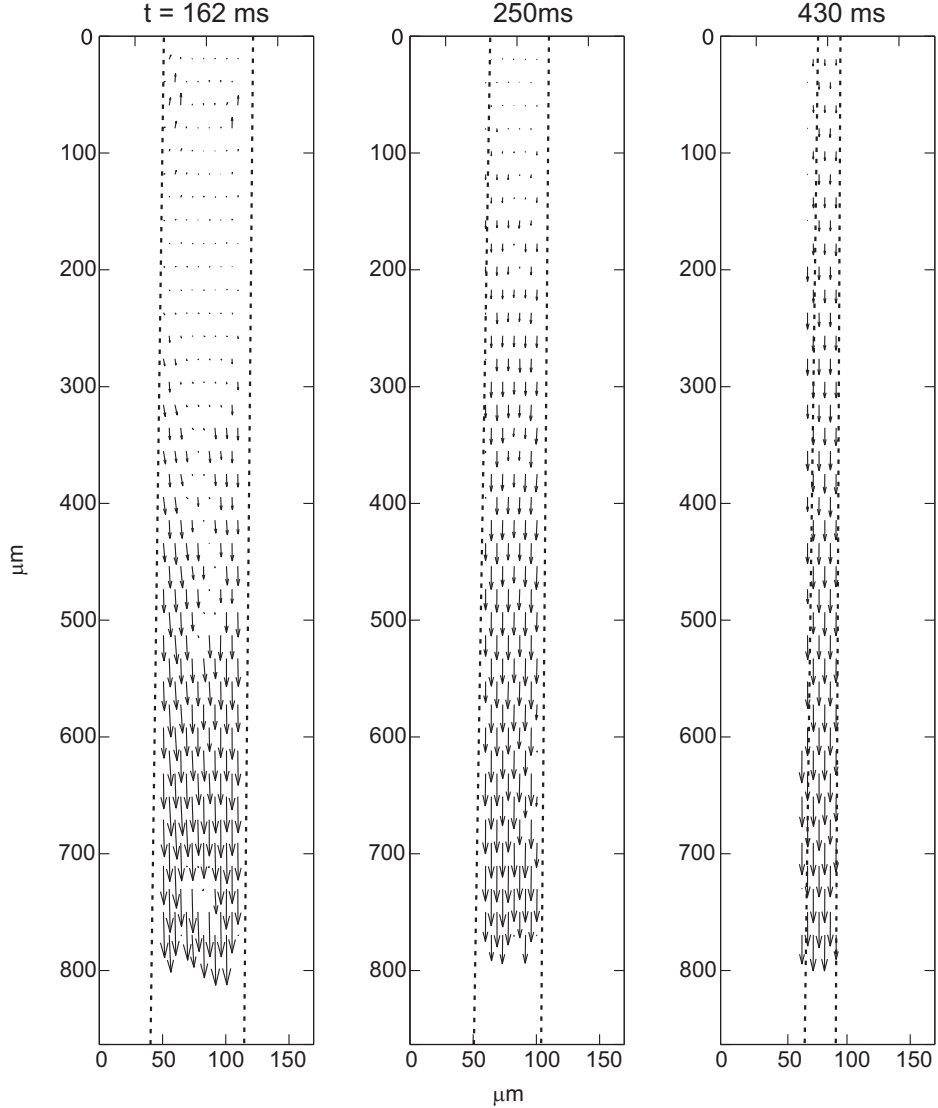}\centering
    \caption{PIV images at three different moments during the capillary breakup (see Fig.\ref{Durchmesser_Kap7}). At the lower part of the filament a fully developed elongational flow is present. Closer to the stagnation point in the upper part of the filament there is stronger axial flow at the boundaries then at the center of the filament. The axial position of the stagnation point moves upward in time. Please remark that the arrows are rescaled from image to image.} \label{PIV-Bilder}
\end{figure}

An overview over the thinning process is given in Fig. \ref{Durchmesser_Kap7}. It shows the filament diameter as a function of time and indicates at which time the data are taken. Figure \ref{PIV-Bilder} shows three PIV-images that are captured at different moments during the capillary breakup of the polymer solution. Obviously, the number of particles strongly decreased in the course of time. Still, the filament was shining fluorescently everywhere because there was not only fluorescence light from the particles in the focal point but also from particles that are out of the focal plane. This effect allowed the determination of the radius $h(t)$ of the filament and verification of whether the plane of measurement was in the middle of the filament. For the first two pictures, only PIV analysis was possible, but for the third picture, both a PIV analysis and a PTV analysis could be performed.

\begin{figure}[h!]
  \includegraphics[width=0.75\linewidth]{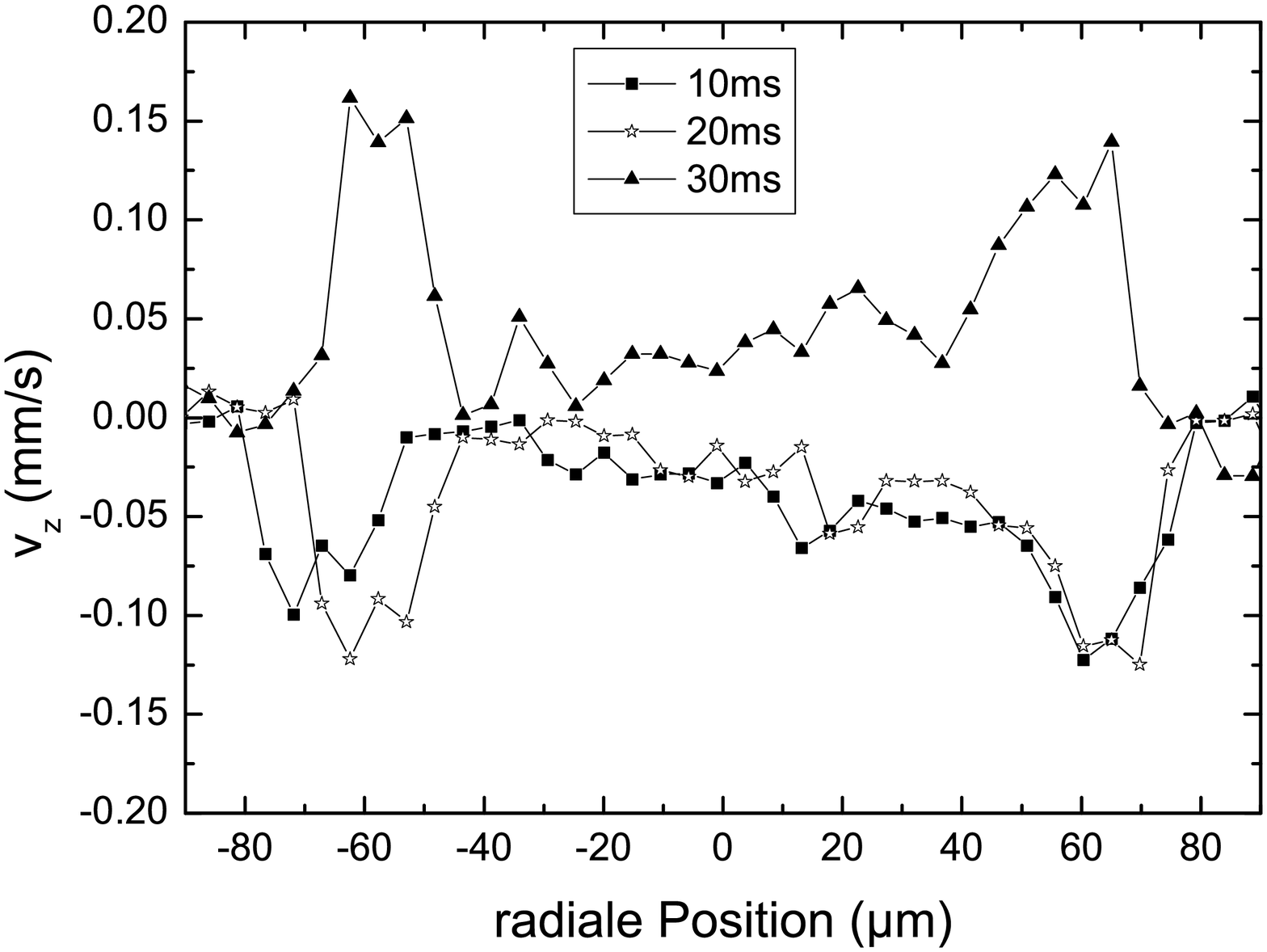}\centering
  \caption{PIV profiles of the axial velocity $v_z$ at the beginning
  of the measurement between t=10ms and 30ms. In this time interval, the filament diameter changes from $h=168$ to $143 \mu m$.} \label{PIV_Anfang}
\end{figure}

 The axial position of the stagnation point of the elongation flow in the first image of Fig. \ref{PIV-Bilder} is at a hight of about $150 \mu m$. Its exact position is difficult to determine, but apparently it was moving upwards and it has vanished from the image section in the third picture. According to Equ. \ref{axial}, the velocity $v_z$ at a fixed axial position should be constant for all times because the elongation rate is a constant. In Fig. \ref{PIV-Bilder}, apparently $v_z$ at a given position is not constant in time. The flow has not yet completely formed in the first images, and the neutral (stagnation) point moves in time. For a more quantitative analysis, a fixed axial position $z$ at $ z = 600 \mu m$ was chosen, and the axial velocity $v_z$ as a function of the radial position $r$ was determined. The result is shown in Fig. \ref{PIV_Anfang} for three profiles at the beginning of the breakup. Positive values mean that the flow goes downwards, whereas negative values represent an upward flow. The three profiles show that there is only a flow near the edge of the filament but no flow in the middle. For t=10 ms and t=20 ms, the liquid flows upwards, whereas, at t=30ms, it flows downwards. The axial velocities are still small, but the flow has changed its direction between 20 and 30ms after the beginning of the measurement. This is because the neutral (stagnation) point has moved upwards as well, and the stationary measurement level is first below and then above the stagnation point.

An inhomogeneity in the flow profile along the r-axes could be explained by at least two different mechanisms. First, the amount of stretch at the boundary is largest due to the larger deformations compared to the center of the filament. This should lead to a stress layer near the boundary. This seems to be indeed the case in some numerical simulations \cite{Harlen2010}. Another possible explanation might be that the flow close to the center and the stagnation point is closer to a pure elongational flow that is more efficient in stretching the polymers and causes a larger elastic stress that suppresses any flow. This could be closely related to what is known from stagnation point flow and is referred to as the "birefringent strand" \cite{Feng97}. It refers to a region of high birefringence (and thus high elastic stress) along the center line close to a stagnation point.

\begin{figure}[h!]
  \includegraphics[width=0.75\linewidth]{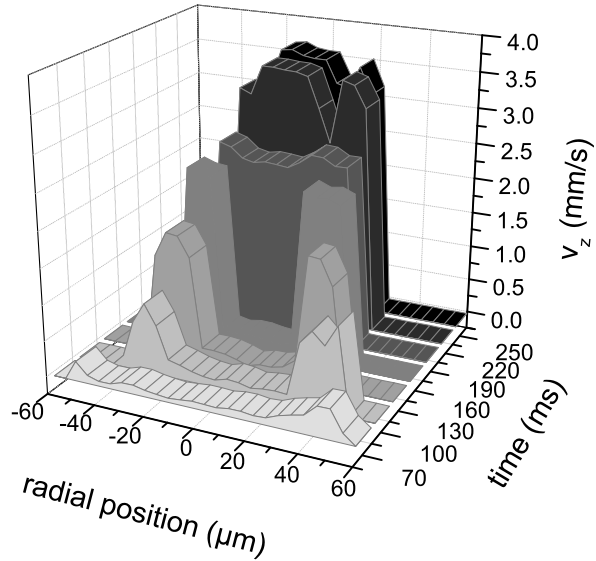}\centering
  \caption{PIV profiles of the axial velocity $v_z$ between 10ms and
  250ms with a time-lag of 30ms. In this time interval the thickness of the filament changes from $h=168 \mu m$ to $48 \mu m$.} \label{PIV-Profile_gesamt3D}
\end{figure}
\begin{figure}[h!]
  \includegraphics[width=0.75\linewidth]{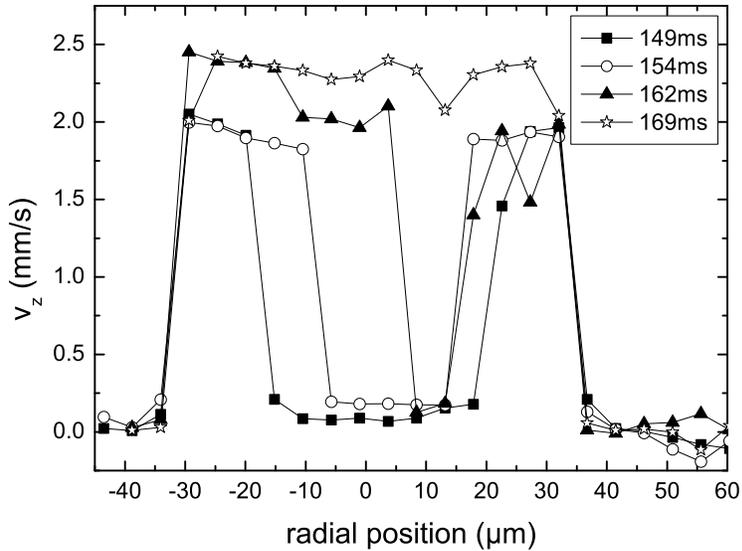}\centering
  \caption{PIV profiles of the axial velocity $v_z$ between 150 ms
  and 170 ms that show the transition from the pure edge flow to the
  plug flow-like profile.} \label{PIV-Profile_Ausschnitt}
\end{figure}

With increasing time, this profile changes qualitatively (Fig. \ref{PIV-Profile_gesamt3D}). The maximum of the axial velocity increases from 0.1 mm/s up to about 3.5mm/s and then remains constant. Moreover, the spatial inhomogeneous profile with some flow at the edge and no flow in the middle of the filament that has formed at the beginning (cf. Fig. \ref{PIV_Anfang}) is conserved only for some time. First, the profile grows until the velocity at the edge has reached a value of about 2.5 mm/s. Then, the expected flow with a nearly constant velocity for all radial positions at a fixed axial position is formed such that the flow "grows" continuously into the center of the filament (Fig. \ref{PIV-Profile_Ausschnitt}). This growth occurs in a manner that is not fully symmetric, but rather growth from one side is preferred. The change from a zero velocity to the finite maximum value at a position $r$ next to the transition from no-flow to flow occurs in a short time interval, i.e., within the 6.5 ms measurement resolution. This could compared to the time scale of viscous  diffusion, which is even much shorter considering the high extensional viscosity present here $\tau= h^2/(\eta_e/\rho)\approx 1\mu s$. The complete transition from the inhomogeneous profile lasts approximately 20 ms after the boundary velocity has reached its maximum at $t=170 ms$ after the beginning of the measurement.

\begin{figure}[h!]
  \includegraphics[width=0.75\linewidth]{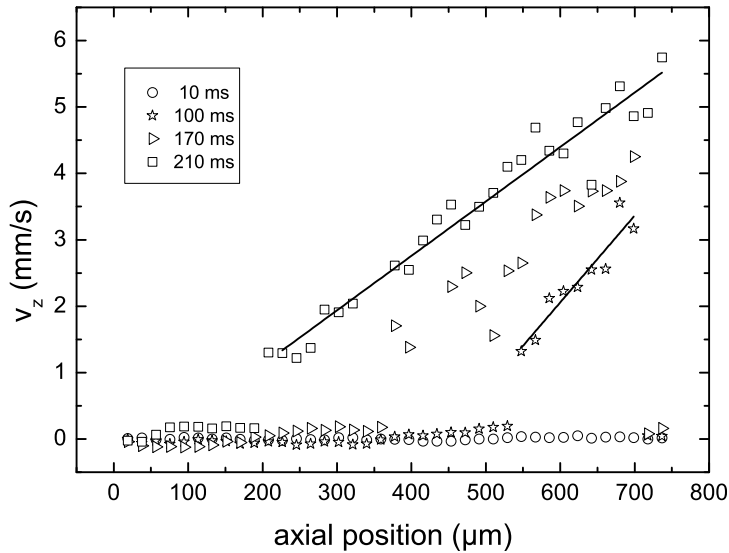}\centering
  \caption{Velocity along the centerline for different times. The elongational flow implies a linear dependency of the axial velocity on the axial position. The slope of the linear fit give the elongational rate $\dot\epsilon$ (see Fig. \ref{Mitte_Dehnungsrate}). Close to the stagnation point, there is a region without any flow and this region moves upward in time (see Fig. \ref{positionneutralpoint}.)  } \label{velocitycenterline}
\end{figure}

\begin{figure}[h!]
  \includegraphics[width=0.75\linewidth]{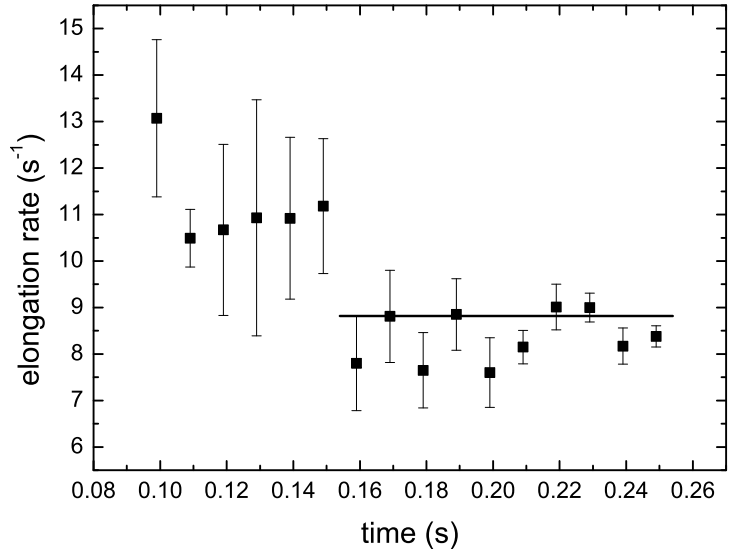}\centering
  \caption{Elongation rate at the center of the filament. Squares
  are calculated by linearly fitting the experimental PIV data of $v_z$ (Fig. \ref{velocitycenterline})
  with respect to $z$ according to equation \ref{axial} . The straight
  line marks the value of $\dot{\varepsilon}$ that is obtained from
  the exponential fit of the diameter (Fig. \ref{Durchmesser_Kap7}).}\label{Mitte_Dehnungsrate}
\end{figure}

\begin{figure}[h!]
  \includegraphics[width=0.75\linewidth]{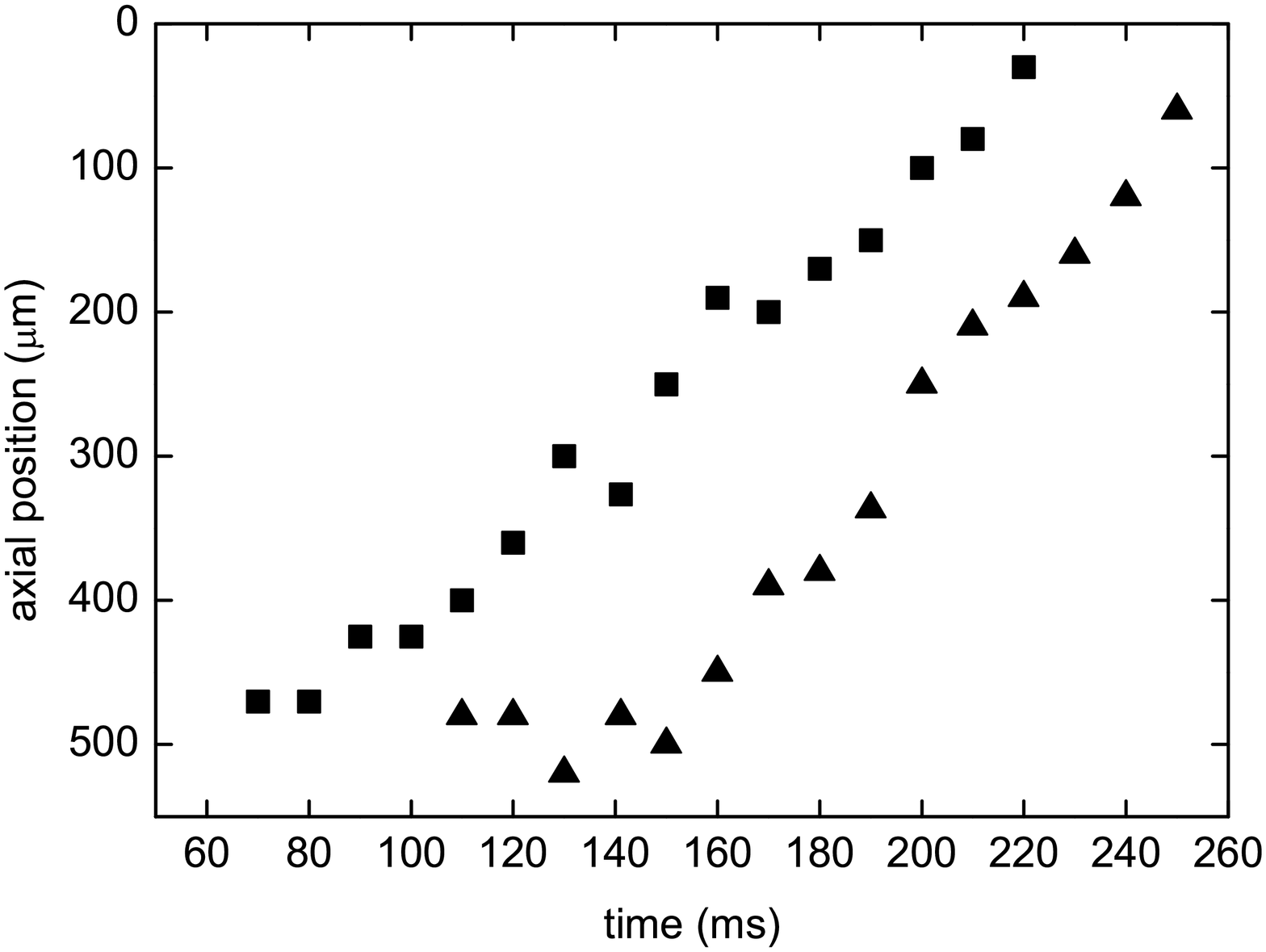}\centering
  \caption{Position of the stagnation point (squares) and position of the region along the centerline with the transition from the elongational flow to zero flow (triangles, see Fig.\ref{velocitycenterline}).} \label{positionneutralpoint}
\end{figure}

If we evaluate the velocities along the filament (Fig. \ref{velocitycenterline}) we see that there is a region without flow and only at some distance from the stagnation point there is flow. The velocity profile in Fig. \ref{velocitycenterline} already suggests that the flow is purely elongational in this regime. The velocity  $v_z$ along the vertical axis can be fitted with the linear law Equ. \ref{axial}. The slope of this linear fit directly gives $\dot{\varepsilon}$. Doing so for the central layer of the filament (parallel to the z-axis) for different times during the thinning process, one gets the results shown in Fig. \ref{Mitte_Dehnungsrate}. The straight line marks the value of $\dot{\varepsilon}$ that is obtained from the relaxation time $\lambda_C$ (cf. Fig. \ref{Durchmesser_Kap7}) that can be deduced from the measurement of the filament diameter $h(t)$ by use of Equ. \ref{Dehnrate2}. There is a sharp transition at $t = 160 ms$. We do not have an explanation for this sharp transition, but this is roughly the moment when there is flow in most part of the filament.  The data at times $t<160 ms$ are taken when the flow profile all over the filament is still inhomogeneous; in principle, one can not yet define a constant elongational rate over the whole filament. For a typical CaBER application, the exponential fit of the diameter $h(t)$ is the most important quantity, and apparently one gets reasonable results if one chooses the range over which the diameter $h(t)$ thins exponentially by visual inspection (here we choose $100 ms < t < 500 ms$). If, for example, one chooses the starting point of the time interval by considering the analysis above and shifts the end for the same amount such that $170 ms < t < 430 ms$, the characteristic time scale changes only by $0.5\%$.

In Fig. \ref{positionneutralpoint} we plot the position of the neutral point and the position of the transition regime from no flow to flow along the centerline. The position of the neutral point is not defined a priori and in principle it can move freely. Though this implies acceleration of fluid elements, the stagnation point indeed moves upward to the upper halfdroplet. With this movement, the transition regime of no flow to flow moves upwards too and therefore the fully developed elongational flow evolves only gradually at a given position.

\subsection{PTV analysis}
\begin{figure}[h!]
  \includegraphics[width=0.75\linewidth]{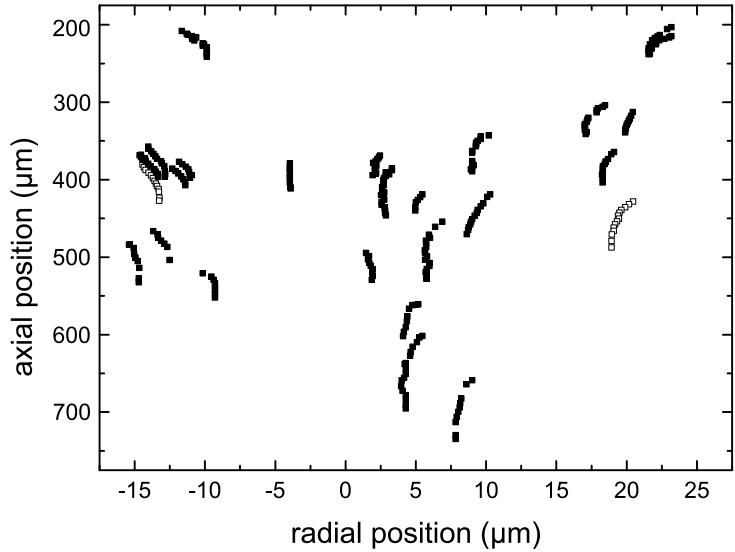}\centering
  \caption{Trajectories of different single particles determined by the PTV analysis. The open symbols are the trajectories that are shown in Fig. \ref{PTV_einzeln}.} \label{PTV_alle}
\end{figure}

\begin{figure}[h!]
  \includegraphics[width=1\linewidth]{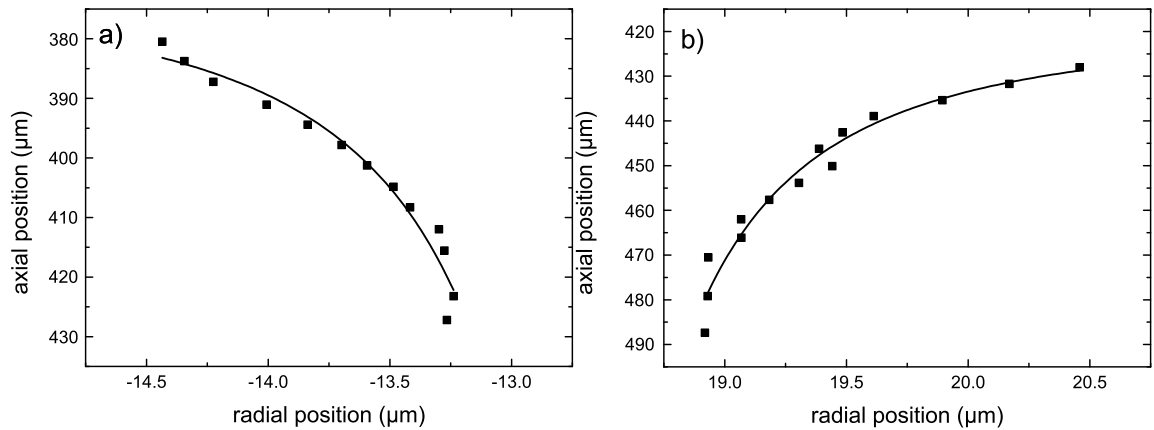}\centering
  \caption{Trajectories of a particle on the a) left-hand side and on the b) right-hand side of
  the filament determined by the PTV analysis. The solid line is a fit according to Equ. \ref{Dehnungsprofil_FIT}.} \label{PTV_einzeln}
\end{figure}

From $t=160 ms$ the flow should be purely elongational and the velocity $\overrightarrow{v}$ has both an axial and a radial component. We tried to deduce the radial component $v_r$ from our PIV data, but $v_r \ll v_z$, and we were not able to resolve $v_r$ in a quantitative manner. In Fig. \ref{PIV-Bilder}, the arrows are nearly parallel to the z-axis, and finally, we had to use a PTV analysis to quantify both velocity components.
The PTV analysis could be used only at longer times when the number of tracer particles became small enough. Furthermore, we wanted to verify if the streamlines are in accordance with the assumption of a pure elongational flow. The result is shown in Fig. \ref{PTV_alle} and \ref{PTV_einzeln}. Resolving Equ. \ref{axial} and \ref{radial} one finds by integration:

\begin{equation}
z=\frac{z_0r_0^2}{r^2} \label{Dehnungsprofil}
\end{equation}

where $z_0$ and $r_0$ are the initial values in the axial and radial direction, respectively. However, in our coordinate system the trajectories do not start at ($r\rightarrow\pm\infty$,$z\rightarrow 0$) and end at ($r\rightarrow 0$,$z\rightarrow\pm\infty$), and the experimental trajectories are given by

\begin{equation}
z=z_0+\frac{a}{(r-r_{\infty})^2}. \label{Dehnungsprofil_FIT}
\end{equation}

$z_0$ again is the initial value in the axial direction (for $r\rightarrow\pm\infty$), and $r_{\infty}$ is the final value in radial direction (for $z\rightarrow\pm\infty$). The latter value is zero for particles on the left side of the filament and positive for those on the right side. The factor a is a free fit parameter that is always positive because $z_0$ is the minimum value in axial direction, and $(r-r_{\infty})^2>0$.

Both trajectories of a particle on the left- and right-hand sides of the thinning filament are shown in Fig. \ref{PTV_einzeln}, and a fit according to Equ. \ref{Dehnungsprofil_FIT} matches the experimental data very well.

\section{Conclusion} \label{sec5}

In conclusion, we presented PIV and PTV data of a thinning capillary bridge of a polymer solution. The optical aberrations that are introduced by the cylindrical filament could be corrected for, and the thinning process could be characterized for a filament thickness reaching from $30 \mu m < h(t) < 150 \mu m$. The measurements at the beginning of the breakup that are analyzed via PIV show that the axial velocity $v_z$ is first strongly varying. Initially, the liquid has a finite velocity only near the edge of the filament, whereas there is no flow in the middle. A possible explanation could be that, before the formation of the filament, there is a stagnation point in the center of the capillary bridge and the stretching and build-up of stresses are most efficient close to the center. Both the slight asymmetric flow profile apparent in Fig. \ref{PIV-Profile_gesamt3D} as well as the non stationarity of the stagnation point might be caused by gravitational sagging. The Bond number $Bo=\rho g h^2/\sigma\approx 10^{-3}$ is very small, but former studies on the final disintegration process of the viscoelastic thread showed that gradational effects still play a role e.g. for the formation of the differently shaped transition regions from the filament to the upper and lower half droplet at the ends of the filament \cite{Gier11}.

 Later on, the inhomogeneous profile evolves into a homogenous one, and the elongation rates (at a fixed axial position) are constant over time for each radial position. The radial velocity $v_r$ was found to be much smaller than $v_z$, and it was not possible to characterize $v_r$ quantitatively with PIV. Therefore, we used PTV to extract the flow lines, and we were able to show that the flow, once fully developed, indeed is a pure elongational one. This happens approximately at the moment when an exponential fit of the filament diameter $h(t)$ becomes reasonable. This quantity is used in simple CaBER (capillary breakup extensional rheology) experiments to deduce the elongational rate and more importantly the elongational viscosity. For such a type of material characterization, we could show that the flow indeed is purely elongational, at least for our polymeric system.

\section*{Acknowledgments}
This work was  supported by the DFG-project WA1336/4 (droplet detachment of complex liquids). We thank Anke Lindner for helpful discussions.


\begin{thebibliography} {10}

\bibitem{E97}
J. Eggers, "Nonlinear dynamics and breakup of free-surface flows", Rev. Mod. Phys. {\bf 69}, 865 (1997).

\bibitem{Wagner2005}
C. Wagner, Y. Amarouchene, D. Bonn, J. Eggers, "Droplet Detachment
and Satellite Bead Formation in Viscoelastic Fluids", Phys. Rev.
Lett. {\bf 95}, 164504 (2005).

\bibitem{Bazilevskii1981}
A.V. Bazilevskii, V.M. Entov, and A.N. Rozhkov, "Orientational effects in the decomposition of streams and strands of diluted polymer solutions", Sov. Phys. Dokl. {\bf 26}, 333 (1981).

\bibitem{Renardy94}
M.~Renardy, "Some comments on the surface-tension driven break-up(or the lack of it) of viscoelastic jets, "J. Non-Newtonian Fluid Mech. {\bf 51} 97 (1994). M.~Renardy, "A numerical study of the asymptotic evolution and breakup of Newtonian and viscoelastic jets", J. Non-Newtonian Fluid Mech. 267 {\bf 59} (1995).

\bibitem{Chang1999}
H.C. Chang and E.A. Demekhin, "Coalescence Cascade towards Drop Formation", J. Fluid Mech. {\bf 380}, 233 (1999).

\bibitem{Stelter2000}
M.~Stelter, G.~Brenn, A.~L.~Yarin, R.~P.~Simgh, and F.~Durst, "Validation and application of a novel elongational device for polymer solutions," J. Rheol. {\bf 44} 595 (2000) . M.~Stelter, G.~Brenn, A.~L.~Yarin, R.~P.~Simgh, F.~Durst, "Investigation of the elongational behavior of polymer solutions by means of an elongational rheometer," J. Rheol. {\bf 46} 507 (2002.

\bibitem{Anna2001}
S.L.Anna, G.H.McKinley, "Elasto-capillary thinning and breakup of model elastic liquids", J Rheol. {\bf 45},115 (2001).

\bibitem{Li2003}
J. Li and M.A. Fontelos, "Drop dynamics on the beads-on-string structure for viscoelastic jets: A numerical study", Phys. Fluids {\bf 15}, 922 (2003).

\bibitem{Etienne2006}
J. Etienne, E.J. Hinch, J. Li, "A Lagrangian-Eulerian approach for the numerical simulation of free-surface flow of a viscoelastic material",  J. Non-Newtonian Fluid Mech.
{\bf 136}, 157 (2006)

\bibitem{Clasen06}
C.~Clasen, J.~Plog, W.~Kulicke, M.~Owens, C.~Macosko, L.~Scriven, M.~Verani, and G.~McKinley, `"How dilute are dilute solutions in extensional flows?,`" J. Rheol. {\bf 50} (2006) 849.
C. Clasen, J. Eggers, M.A. Fontelos, J. Li, and G.H. McKinley, "The beads-on-string structure of viscoelastic threads," J. Fluid Mech. {\bf 556}, 283 (2006).

\bibitem{Tirtaatmadja06}
V. Tirtaatmadja, G. H. McKinley, J. J. Cooper-White, "Drop formation and breakup of low viscosity elastic fluids: Effects of molecular weight and concentration", Phys. Fluids {\bf 18}, 043101 (2006).


\bibitem{Morrision2010}
N.F. Morrison and O. G. Harlen, "Viscoelasticity in Inkjet Printing", Rheol. Acta {\bf 49}, 1435, (2010)

\bibitem{Zell2010}
A. Zell, S. Gier, S. Rafa\"{\i}, and C. Wagner, "Is there a relation between the relaxation time measured in CaBER experiments and the first normal stress coefficient?", J. Non-Netwonian Fluid Mech. {\bf 165}, 1265 (2010).

\bibitem{Bhat2010}
P.P. Bhat, S. Appathurai, M.T. Harris, M. Pasquali, G.H. McKinley, O.A. Basaran, "Formation of beads-on-a-string structures during break-up of viscoelastic filaments", Nature Phys. DOI: 10.1038/NPHYS1682.

\bibitem{Sattler2008} R. Sattler, C. Wagner, J. Eggers, "Blistering Pattern and Formation of Nanofibers in Capillary Thinning of Polymer Solutions", Phys. Rev. Lett. {\bf 100}, 164502 (2008).

\bibitem{Meinhart99}
C. D. Meinhart, S. T. Wereley, and J. G. Santiago, "PIV measurements of a microchannel flow",  Exp. Fluids {\bf 27}, 414 (1999).

\bibitem{Mielnik04}
M. M. Mielnik, L. R. S${\ae}$tran, "Micro Particle Image velocimetry - an overview", Turbulence {\bf 10}, 83 (2004)

\bibitem{Rothert01}
A. Rothert, R. Richter, I. Rehberg, "Transition from Symmetric to Asymmetric Scaling Function before Drop Pinch-Off", Phys. Rev. Lett.
{\bf 87}, 084501 (2001).

\bibitem{Amarouchene01}
Y. Amarouchene, D. Bonn, J. Meunier, H. Kellay, "Inhibition of the Finite-Time Singularity during Droplet Fission of a Polymeric Fluid", Phys. Rev. Lett. {\bf 86}, 3558 (2001).

\bibitem{CooperWhite02}
J. J. Cooper-White, J. E. Fagan, V. Tirtaatmadja, D. R. Lester, D. V. Boger, ``Drop formation of constant low-viscosity, elastic fluids", J. non-Newtonian Fluid Mech. {\bf 106}, 29 (2002).

\bibitem{Lindner03}
A. Lindner, J. Vermant, D. Bonn, "How to obtain the elongational viscosity of dilute polymer solutions?", Physica A {\bf 319}, 125 (2003).

\bibitem{Rodd05}
L. E. Rodd, T. P. Scott, J. J. Cooper-White, G. H. McKinley, "Capillary Break-up Rheometry of Low-Viscosity Elastic Fluids", Appl. Rheol. {\bf 15}, 12 (2005).

\bibitem{Yao00}
M. Yao, S.H. Spiegelberg,G. H. McKinley, "Fluid Dynamics of Weakly Strain-Hardening Fluids in Filament Stretching Devices", J. Non-Newt. Fluid Mech., {\bf 89},1 (2000).

\bibitem{McKinley01}
G.H. McKinley, O. Brauner, M. Yao, "Kinematics of Filament Stretching in Dilute and Concentrated Polymer Solutions", Korea-Australia Rheol. J. {\bf 13}, 29 (2001).

\bibitem{Rothstein02}
J.P. Rothstein and G.H. McKinley, "A comparison of the stress and birefringence growth of dilute, semi-dilute and concentrated polymer solutions in uniaxial extensional flows", J. Non-Newtonian Fluid Mech., {\bf 108} 275 (2002).

\bibitem{Miller09}
E.~Miller, C.~Clasen, and J.~P. Rothstein, "The effect of step-stretch parameters on capillary breakup extensional rheology (CaBER) measurements", Rheol. Acta {\bf 48} 625, (2009).

\bibitem{Devasenathipathy03}
S. Devasenathipathy, J. G. Santiago, S. T. Wereley, C. D. Meinhart, K. Takehara, "Particle imaging techniques for microfabricated
fluidic systems", Experiments in Fluids \textbf{34}, 504-514 (2003).

\bibitem{MatPIV}
J. K. Sveen, {\it{An introduction to MatPIV v. 1.6.1}} (Dept. of Math. University of Oslo, "Mechanics and Applied Mathematics", NO. 2 ISSN 0809-4403, August 2004).

\bibitem{Harlen2010}
O. Harlen, University of Leeds, priv. comm..

\bibitem{Feng97}
J. Feng and L.G. Leal, "Numerical simulations of the flow of dilute polymer solutions in a four-roll mill", J. Non-Newt.
Fluid Mech.  {\bf 72} 187 (1997).

\bibitem{Gier11}
S. Gier, R. Sattler, J. Eggers, C. Wagner, "The final stages of capillary break-up of polymer solutions", Phys. Fluids {\bf 24} 023101 (2012).

%
%
%
%
%
%
%
%
%
%
%
%
%
%
%
%
%
%
%

\end{thebibliography}

\end{document}